\documentclass[runningheads]{llncs}
\usepackage{graphicx}
\usepackage{xcolor}
\usepackage{listings}
\usepackage{calc}

\usepackage{todonotes}

\usepackage[acronym]{glossaries}
\newacronym{cuda}{{CUDA}}{compute unified device architecture}
\newacronym{hip}{{HIP}}{Heterogeneous-compute Interface for Portability}
\newacronym{gpgpu}{GPGPU}{general purpose computing on GPUs}
\newacronym{tbb}{TBB}{Thread Building Blocks}
\newacronym{cas}{CAS}{atomic compare and swap}

\newacronym{jit}{{JIT}}{just-in-time compilation}
\newacronym{tmp}{TMP}{template metaprogramming}
\newacronym{raii}{RAII}{resource acquisition is initialization}
\newacronym{dsl}{DSL}{domain-specific language}
\newacronym{stl}{STL}{standard template library}
\newacronym{ice}{ICE}{internal compiler error}

\newacronym{pic}{PIC}{particle-{}in-{}cell}

\begin{document}
\title{Challenges Porting a C++ Template-Metaprogramming Abstraction Layer to
Directive-based Offloading}
\titlerunning{Porting a C++ Abstraction Layer to Directive-based Offloading}
%
\author{Jeffrey~Kelling\inst{1}\orcidID{0000-0003-1761-2591} \and
Sergei~Bastrakov\inst{2}\orcidID{0000-0003-3396-6154} \and
Alexander~Debus\inst{2}\orcidID{0000-0002-3844-3697} \and
Thomas~Kluge\inst{2}\orcidID{0000-0003-4861-5584} \and
Matt~Leinhauser\inst{3,4}\orcidID{0000-0003-2914-1483} \and
Richard~Pausch\inst{2}\orcidID{0000-0001-7990-9564} \and
Klaus~Steiniger\inst{2}\orcidID{0000-0001-8965-1149} \and
Jan~Stephan\inst{4}\orcidID{0000-0001-7839-4386} \and
Ren\'e~Widera\inst{2}\orcidID{0000-0003-1642-0459} \and
Jeff~Young\inst{4,5}\orcidID{0000-0001-9841-4057} \and
Michael~Bussmann\inst{4}\orcidID{0000-0002-8258-3881} \and
Sunita~Chandrasekaran\inst{3}\orcidID{0000-0002-3560-9428} \and
Guido~Juckeland\inst{1}\orcidID{0000-0002-9935-4428}}
\authorrunning{J. Kelling et al.}
%
\institute{Department of Information Services and Computing,
Helmholtz-Zentrum Dresden-Rossendorf (HZDR), Bautzner Landstr. 400, 01328 Dresden, Germany \and
Insitute of Radiation Physics,
Helmholtz-Zentrum Dresden-Rossendorf (HZDR), \\Bautzner Landstr. 400, 01328 Dresden, Germany \and
Deptartment of CIS, University of Delaware, Newark, Delaware, 19716, USA \and
Center for Advance Systems Understanding (CASUS), Am Untermarkt 20
02826 G\"orlitz, Germany \and
Georgia Tech, School of Computer Science, Atlanta, GA 30332, USA
}
\maketitle              

\setcounter{footnote}{0}
\renewcommand{\arraystretch}{1.4}
\begin{abstract}
 HPC systems employ a growing variety of compute accelerators with different
 architectures and from different vendors. Large scientific applications 
 are required to run efficiently across these systems but need to retain a
 single code-base in order to not stifle development. Directive-based
 offloading programming models set out to provide the required portability,
 but, to existing codes, they themselves represent yet another API to port to.
 Here, we present our approach of porting the GPU-accelerated particle-in-cell
 code PIConGPU to OpenACC and OpenMP target by adding two new backends to its
 existing C++-\glsdesc{tmp}-based offloading abstraction layer alpaka and
 avoiding other modifications to the application code. We introduce our approach
 in the face of conflicts between requirements and available features in the
 standards as well as practical hurdles posed by immature compiler support.

\keywords{C++ \and OpenACC \and OpenMP \and Offloading.}
\end{abstract}

\section{Introduction\label{s:intro}}

Contemporary scientific applications are often written with accelerators, like
GPUs, in mind. This approach is required to make use of many modern compute
clusters and supercomputers, but it brings a dilemma of choice from a zoo of
proprietary and open offloading APIs coming with varying degrees of support form hardware
vendors. While vendor specific, proprietary APIs usually promise best performance and
hardware support, their use limits an application to one vendor's ecosystem.
Open APIs offer portability, but cannot guarantee support on future hardware.
Committing to any particular offloading API may thus necessitate future rewrites
of software when new hardware or architectures become available or the chosen
API looses support, which may be unfeasible with the limited time-budged for
software development available within the scientific community. Porting even
between similar offloading APIs has in the past also proven to be a non-trivial
task~\cite{JuckelHernanJacobNeilso2016_DescribiPrescrib}.

With C++ being very prevalent in contemporary HPC, one approach to mitigate this
problem of choice is to use this language's strength in zero-overhead abstraction
via templates and \gls{tmp} to create an abstraction layer which applications
can use to formulate parallelism in an API agnostic way. At first glance this only
moves the support problem from the offloading API to the abstraction layer,
however, in relation to a large scientific application such an abstraction layer
is much smaller and thus requires much less work to port to new parallel APIs.
Thus, such an abstraction layer is first conceived as a part of one or a group
of scientific applications rather than a library for a general audience. Despite
this focus, it will lend itself to other applications, with support being
guaranteed as long as the primary applications fuel its developer's interest.

Another breed of open-standard APIs for parallelization and offloading are
directive-based approaches which try to minimize changes to an existing, sequential
code base by employing directives for marking code regions which may be offloaded by the
compiler without modifying the code in it base-language to the eyes of a
compiler with support not enabled. Like with any other API, their practical
portability depends on the development afforded by compiler and hardware
vendors, too. Currently two competing directive-based model that target accelerator offloading are being developed:
OpenMP~\cite{openmp}, which has be around since 1997 as a parallel model for multi-core CPUs,
has been extended with \texttt{target} directives in version 4.0 of its standard, while
OpenACC~\cite{openacc} was initially created to provide offloading to accelerators exclusively.

This paper presents our efforts and experiences in porting the computational
radiation physics code PIConGPU~\cite{PIConGPU2013} to both OpenACC and OpenMP
\texttt{target}.  PIConGPU uses alpaka~\cite{ZenkerAsHES2016}, which was first
developed to provide portable between different accelerator architectures,
including GPUs and Intel MIC, as well as multi-core CPUs to PIConGPU.  Even
though the primary purpose of OpenMP and OpenACC is to simplify porting of
existing codes, porting a code like PIConGPU, with alpaka, as a C++ abstraction
layer, already in place, the code for two new alpaka back-ends requires much less
effort than porting the whole of PIConGPU directly.  Thus, the majority of this
paper is dedicated to documenting our efforts to create alpaka back-ends for both OpenACC and
OpenMP \texttt{target}. Some of the lessons learned here will also apply to
using directives in C++ codes in general and specifically other C++ offloading
abstraction layers.

PIConGPU is an extremely scalable, heterogeneous, fully relativistic
\gls{pic} C++ code. This code has been chosen as one of the eight
CAAR codes across the United States by the Oak Ridge National Laboratory. CAAR
stands for Center for Acceleration Application Readiness - a program at ORNL
that is created to ready applications for their next generation computing
system, Frontier, the first exascale system to be in place later this year,
2021. Work in this paper narrates challenges and potential solutions to prepare
such large scale applications for the ever changing hardware platforms.

This paper discusses the aspects of programmability, portability and per\-for\-mance
via the following contributions:
\begin{itemize}
 \item challenges porting PIConGPU to a directive-based programming model,
 \item creating alpaka back-ends for OpenMP and OpenACC and evaluating the
  functionalities of the available compilers, and
 \item highlighting compiler and runtime issues throughout the PIConGPU code
 migration process.
\end{itemize}

We provide an overview of various offloading APIs and abstraction
libraries in Section~\ref{s:relatedWork}. In Section~\ref{s:methods} the
employed APIs and alpaka are reviewed and compared. During the porting process,
various issues with both standards and compiler support were encountered which
are described in Section~\ref{s:porting}. In Section~\ref{s:issues} we first
discuss the major obstacles we faced during this effort, then provide a short
overview of which examples already work using our OpenMP target and OpenACC
alpaka backends and current compilers. Finally, Section~\ref{s:concl}
concludes with an outlook on the developments in the OpenMP and OpenACC
ecosystems.

\section{Related Work\label{s:relatedWork}}

Since the dawn of \gls{gpgpu}, quite a few offloading APIs have come and gone. One
notable open, cross vendor entry is the Khronos Group's OpenCL~\cite{opencl}
which is barely used anymore in HPC mostly due veining vendor support in this
space and partly due to its choice of having device code separate from the host
source. Other APIs use a \emph{single-source} approach, where offloaded
code is integrated into the host language. NVIDIA's proprietary
\gls{cuda}~\cite{cuda}, which is older than the OpenCL standard,
is a C/C++ dialect which offers single-source offloading. It is the most widely
used API for scientific codes today, having evolved into the go-to API for many
scientific application developers. In response to \gls{cuda}'s popularity, AMD
created \gls{hip}, which mimics the \gls{cuda} API to simplify porting \gls{cuda} codes.
Later single-source approaches aimed at integrating offloading into standard C++
without creating a dialect. This was first attempted by Microsoft introducing
C++ AMP~\cite{amp}, which inspired Khronos' SYCL~\cite{sycl}.
The \texttt{do concurrent} construct introduced to Fortran in version 2008,
which can also be offloaded to GPUs by some compilers~\cite{nvhpcDoConcurrent},
is a more mature example of the drive to include offloading
support directly into base languages.

With these and more offloading APIs, including OpenMP \texttt{target} and
OpenACC, available to chose from obsolescence is always looming and can not only
be caused by a vendor dropping support for an API, but also by technical
developments: With the prevalence NVIDIA GPUs in supercomputers, \gls{cuda} looked
like a safe choice not long ago. A switch to an AMD-based machine
could be addressed by an, in theory, not too demanding port to \gls{hip}. However,
both of these APIs can only target GPUs not the host CPUs which are the primary
compute resource installed on the currently fastest supercomputer
Fugaku~\cite{top500fugaku}, meaning a code based on these APIs would not be able
to run in such a system. Therefore, C++ abstraction layers do not only aim
to be a bridge between accelerator architectures and APIs, but also include
sequential and parallel execution on the host.

Sandia National Laboratories started the development of
Kokkos~\cite{CarterEdwards20143202} to achieve performance portability for
scientific codes written in C++. It currently supports \gls{cuda}, OpenMP (host),
native POSIX threads and HPX. Another library is RAJA~\cite{raja}, developed at
Lawrence Livermore National Laboratory, which offers support of \gls{cuda},
\gls{hip}, \gls{tbb} and OpenMP (including \texttt{target}). RAJA also
specifically aims to provide parallel algorithms like Thrust~\cite{thrust}.
These libraries, just like alpaka, are single-source, passing user code as a
predicate to templates which encapsulate the underlying back-end APIs. In
contrast to these examples, alpaka aims to be lightweight low-level abstraction,
providing generic access to the basic parallel hierarchies of accelerators
without shaping user code prescribing memory layout or
algorithms~\cite{MattheWideraZenkerBussma2017_TuningAndOptimization}.

Another
approach that should be mentioned is that of a template expression library, where
user code is not directly passed to a back-end but written in \gls{tmp}-based
domain specific language which then generates backend-compatible code.
VexCL~\cite{DemidoAhnertRuppGottsc2013_ProgrammingCudaAndOpencl} used this
approach to generate code for \gls{jit} to support OpenCL, next to \gls{cuda} and
OpenMP.

%

\section{Methods and APIs\label{s:methods}}

\subsection{Alpaka and PIConGPU}

\paragraph{PIConGPU} is a plasma physics code that simulates the dynamics of
fast charged particles  in  electromagnetic  fields  taking  into  account
relativistic  effects  and fields  generated  by  moving  particles.  The
employed  particle-in-cell  approach does  store  electric  and  magnetic
fields  on  a  regular  grid  but  also  must  keep track of quickly moving
particles.  Particles are stored on a per-supercell basis using a dynamic
collection of fixed-sized particle vectors.  Kernels implementing particle-grid
operations on GPUs collectively copy nearby field values between device memory
and on-chip scratch memory (e.g. \verb|__shared__| in CUDA) to facilitate
efficient parallel processing of particles.  This approach takes the
hierarchical architecture of GPUs into account and was initially implemented
using CUDA directly.

PIConGPU uses templates and \gls{tmp} to
provide a \gls{dsl} for the user to describe the types
of particles, interactions and external fields present in the simulation. Some
numerical parameters of these are also fixed at compile time. This incurs a
large number of \texttt{constexpr} variables of static lifetime, either as
static class members or global inside namespaces, to be defined as part of the
simulation definition. 

\paragraph{Alpaka} was introduced to make PIConGPU portable without duplicating
code for different target architectures while retaining the present performance
on GPUs. To this end, alpaka provides abstract forms of all major concepts
present in CUDA, including the execution hierarchy, atomics, block-level synchronization
primitives and shared memory as well as a generalize device and memory
management API. To generalize these concepts alpaka provides a rather verbose
API. To simplify porting codes from CUDA to alpaka the wrapper library
cupla~\cite{cupla} was created which provides a simplified API which is more
similar to CUDA. Other applications also adopted alpaka as their abstraction for
offloading, such as a ptychography code and a high-level trigger software in a
large-scale particle detector.

Alpaka currently provides a backend targeting sequential execution, backends for parallel
execution on the host via OpenMP and \gls{tbb}, as well as on GPUs using CUDA and
HIP. A SYCL backend is also under development~\cite{alpakaSYCL}.
Alpaka's codebase~\cite{alpaka} contains some examples, ranging from a parallel
``Hello World'' to Euler integration. All backends are also covered by a suite
of tests, each of which tests a different aspect of offloading. One test
focuses on transferring memory between buffers on host and device, including
multiple cases covering different source and device locations and buffer sizes.
Other tests cover aspects of kernel execution on the device, like trivial
kernels, variables in block-shared memory, synchronization between threads or
atomic operations.

\subsection{Review of OpenACC and OpenMP target}

Both OpenACC and OpenMP \texttt{target} have been designed with GPUs or similar
accelerators as offload target devices in mind. Both expose the same two main layers of
parallelization on target devices as known from GPUs: A work grid is first decomposed into
loosely coupled blocks executing independently, which in turn decompose into
threads which execute in at least such a way that data sharing may be
exclusively between threads of the same block. Both models can also target CPUs,
where implementations tend to limit the number of threads per block, often
to one, in order to effectively map blocks to CPU threads.
An overview on the names the
different models assign to these layers is provided in
table~\ref{tab:alpakaOmpOaccLayers}. We will stick to the CUDA/Alpaka naming
scheme.

\begin{table}[bt]
 \centering
 \caption{\label{tab:alpakaOmpOaccLayers}
  Rosetta-stone for the names of parallel execution layers. The
  \texttt{element}/\texttt{simd} layer is only listed for completeness.
  Each model using a different set of terms pollutes the name space, which is
  why we stick to the CUDA/Alpaka terminology in the text.
  Where the word ``worker'' is used it refers to a scheduled unit executing some
  work on any parallel level, not to the OpenACC \texttt{worker}.
 }
 \begin{tabular}{llll}
  \multicolumn{1}{c}{CUDA} & \multicolumn{1}{c}{Alpaka} &
  \multicolumn{1}{c}{OpenMP 5.0} & \multicolumn{1}{c}{OpenACC 3.0} \\
  \hline
   grid & grid & (target) & (parallel) \\
   block & block & team & gang \\
   thread & thread & thread & worker \\
   --- & element & simd & (vector) \\
 \end{tabular}
\end{table}

OpenACC adds a third innermost layer called \texttt{vector} as a straight
continuation of this concept, which could conceivably be translated in the same
way as OpenMP's \texttt{simd} loop on some architectures. While alpaka does
provide an \texttt{element} layer inside threads, no abstraction is provided
beyond offering a work-partitioning concept to support canonical \texttt{for}
loops, ideally with compile-time length which a compiler can automatically
vectorize.\footnote{Alpaka is following the approach of very long instruction
word (VLIW) architectures in OpenCL in this aspect.}
As this layer is not going to be mapped explicitly we are not covering it further.

Both models provide directives to control parallel execution and data
movement between host and device. Directives start with
\begin{align*}
  &\texttt{\#pragma acc ...}&&\text{in OpenACC and}\\
  &\texttt{\#pragma omp ...}&&\text{in OpenMP.}
\end{align*}
Here, we shall briefly review the basic
primitives of the OpenACC 3.0~\cite{openacc30} and OpenMP 5.0~\cite{openmp50}
APIs to highlight their primary design aspects that affect our porting
efforts.

Execution of code on the device is initiated using the directives
\texttt{acc parallel} and \texttt{omp target}.
OpenMP provides the directives \texttt{teams} and \texttt{parallel} which cause
regions of code to be executed by multiple blocks or threads, but do not imply
any distribution of work. Loops can be distributed in these regions using the
\texttt{distribute} (blocks) and \texttt{for} (threads)
directives.
In OpenACC parallel execution and work distribution constructs are inseparably
linked with work distribution provided by the \texttt{loop gang} and
\texttt{loop worker} constructs. Herein lies the main conceptual difference
between the OpenMP and OpenACC programming models: While OpenMP aims to support
all types of parallelism, OpenACC is exclusively designed to describe data parallelism in device code.

On GPUs no assumptions about the execution order of blocks should be made, which is
reflected in OpenMP not allowing synchronization and locks across blocks, it
does, however, support them between threads of the same block, using
\texttt{barrier} and \texttt{critical}, respectively. Following its
exclusively data-parallel paradigm, OpenACC does not make any assumptions on
execution order on any level and thus does not support explicit synchronization
and locks. Because a \texttt{loop gang} region may contain multiple loops in
sequence marked with \texttt{loop worker} and \texttt{loop vector}
synchronization can be achieved implicitly within blocks. 

Both APIs support atomic operations through an \texttt{atomic} directive. OpenMP
defines atomics as binding to the target device, OpenACC is less clear on this
aspect, only stating that operations are atomic between gangs.

OpenMP provides the runtime API functions \texttt{omp\_get\_team\_num()} to
determine the id of a block in device code and \texttt{omp\_get\_thread\_num()}
determine the id of a thread within a block. OpenACC does not provide this
functionality because in its strict map-reduce picture only the work items must
be identified, the workers may remain anonymous to the application code.

\subsubsection{Data Management and Declarations}

The canonical way of moving data between host and device in these models is by
declaring host storage locations to be associated to locations in device
memory and later, or in the same construct, instructing the runtime to copy data
between the host and associated device locations. We will not elaborate on the
\texttt{omp target data} and \texttt{acc data} directives because in our port we
are using runtime API functions which allow for explicit memory management
instead. These are listed in table~\ref{tab:memAPI}.

\begin{table}[t]
 \caption{\label{tab:memAPI}
  Overview of OpenMP and OpenACC manual memory management routines with CUDA
  versions listed as reference.
 }
 \centering
  \begin{tabular}{lll}
  \multicolumn{1}{c}{CUDA} &
  \multicolumn{1}{c}{OpenMP 5.0} & \multicolumn{1}{c}{OpenACC 3.0} \\
  \hline
   \texttt{cudaMalloc} & \texttt{omp\_target\_alloc} & \texttt{acc\_malloc} \\
   \texttt{cudaMemcpy} & \texttt{omp\_target\_memcpy} &
    \begin{minipage}[t]{28ex}
     \texttt{acc\_memcpy\_to\_device} \\
     \texttt{acc\_memcpy\_from\_device} \\
     \texttt{acc\_memcpy\_device}
    \end{minipage}\\
   \texttt{cudaFree} & \texttt{omp\_target\_free} & \texttt{acc\_free} \\
 \end{tabular}
\end{table}

Clauses related to data movement may also be present in \texttt{omp target} and
\texttt{acc parallel} constructs where they declare how the listed variables
should be handled before the start and after the end of the attached
region. On the OpenACC side these include \texttt{copy},
\texttt{copyin}, \texttt{copyout}, \texttt{create} and others. OpenMP uses the
\texttt{map} clause, which handles details through a \texttt{map-type} attribute. 

Global variables, including compile-time constants, must be \emph{declared} for
presence on the device. In OpenACC this is done through the construct
\texttt{acc declare copyin( \textit{varList}~)}, where \textit{varList} is a list
variables that are declared for the device. With OpenMP global variables can be
made available to device code either, in a similar way, using the \texttt{omp
declare target( \textit{varList}~)} construct or by declaring them inside a
\texttt{declare target} region.  Implementations relax this requirement and only
require declaration of compile time constants if they are not optimized out and
the compiler places the value in run-time memory. Violations of this relaxed
requirement get expressed as linker errors.


Similar constructs are also used to declare functions to be compiled for the
device, so they can be called from offloaded code in other compilation units.
Functions defined in the same compilation unit can be called without having been
declared for the device.

\subsection{Experimental Setup}

Initially, we started out testing various compilers. For the OpenMP we tried
\emph{Clang}, AMD's Clang-based development compiler AOMP version 0.7, AMD's ROC
variant of Clang version 4.3.0 (based on Clang 13)~\cite{rocmclang}, GCC
version 9, and IBM XL version 16.1.1-5. For
OpenACC NVIDIA's \emph{NVHPC} toolkit versions from 20.3 to 21.7 and GCC
versions 9 through 11 were used. We stopped testing with GCC and
IBM XL due to fixes to reported bugs not becoming available timely enough to fit
the scope of this work. In the case of IBM XL, debugging runtime errors also
turned out to be infeasible due to prohibitively long compile times. We dropped
AOMP due repeated difficulty building new releases and also most fixes to bugs
AOMP inherited from upstream clang being available upstream first.

Extensive testing was done only for OpenMP using Clang version 10 and above,
primarily tracking the \texttt{main} branch,
targeting host (x86\footnote{Intel i7-4930K\label{fn:cpux86}, Ubuntu 18.04}) and AMD GPUs
(HSA\footnote{Radeon Vega 64, Ubuntu 18.04}) and for OpenACC using the latest NVHPC
release, currently 21.7, targeting host (x86\textsuperscript{\ref{fn:cpux86}}) and
NVIDIA GPUs\footnote{GTX Titan Black
, Ubuntu 18.04}.
We only used ROC Clang for a few tests prior to this paper.
Our testing was strictly focussed on
functionality and therefore performance was not measured.

\section{Porting Alpaka\label{s:porting}}

With PIConGPU using alpaka to abstract parallel execution models, the only
viable way of porting it to OpenMP and OpenACC is to create two alpaka backends
and, ideally, not touch PIConGPU itself at all. This severely reduces the amount
of code required for the port because each alpaka feature only needs to be
mapped to the respective target model once, resulting in most relevant OpenMP and
OpenACC construct being used only once in the code.

Alpaka breaks down any offloading backend into a set of basic concepts which are
separately implemented and tied together. Table~\ref{tab:alpakaConcepts}
lists the most relevant of these.

\begin{table}[t]
  \centering
  \caption{\label{tab:alpakaConcepts}%
  List of selected primary concepts which need to be implemented by any alpaka backend.}
  \begin{tabular}{lp{\linewidth/4*3}}
    \texttt{Acc} &
     A type that acts as a handle for a backend. It can be used at compile time
     to select implementations of traits related to the backend and thus ties
     all parts of a backend together.
     \\
    \begin{minipage}[t]{16ex}
      \texttt{AtomicGrids} \\
      \texttt{AtomicBlocks} \\
      \texttt{AtomicThreads}
    \end{minipage} &
     Atomic operations which can be implemented differently for atomicity at
     each level of parallelization: on the whole target device, between block of
     a kernel, between threads of a block.
     \\
    \texttt{BlockSharedDyn} &
     A block-shared memory buffer of a size that is set at run time.
     \\
    \texttt{BlockSharedSt} &
     A strategy to declare block-shared variable akin to CUDA's
     \verb|__shared__|.
     \\
    \texttt{BlockSync} &
     A strategy by which to synchronize threads within a block.
     \\
    \texttt{Buf} &
     A \glsname{raii} class managing a memory buffer in a target device.
     \\
    \texttt{Dev} &
     A handle for a target device.
     \\
    \texttt{IdxBt} &
     A strategy by which a thread can determine its id within a block.
     \\
    \texttt{IdxGb} &
     A strategy by which a block can determine its id within a kernel.
     \\
    \texttt{Queue} &
     A type that implements an execution queue on a target device.
     \\
    \texttt{TaskKernel} &
     A class that wraps user code for execution on a target device.
     \\
    \texttt{WorkDiv} &
     A type holding information about the requested sizes of the grid, blocks
     and threads (number of elements) for kernel execution.
     \\
  \end{tabular}
\end{table}

\subsubsection{Executing User Code on Device}

For each backend, alpaka provides a type \texttt{TaskKernel\texttt{BackendName}}
which wraps a user-provided
functor containing the payload code as well as the arguments that
should be passed to the functor on-device for execution in a \texttt{Queue}.
This is where alpaka's parallel levels get mapped onto the levels provided by the
backend. The listing in Figure~\ref{fig:taskKernel} illustrates the basic structure of what the alpaka
\texttt{TaskKernel} invocation could look like for both OpenMP and
OpenACC.

\begin{figure}[tp]
 \centering
\begin{lstlisting}[language=C++]
template<class Functor, class... Args>
void TaskKernel_OpenMP5_OpenACC (
  WorkDiv workDiv,   // grid size
  Functor functor,   // user functor
  Args ...args )     // user arguments
{

  // OpenMP
# pragma omp target |\label{lst:ompTarget}|
  {
# pragma omp teams distribute
    for ( int blockIdx = 0;
      blockIdx < workDiv.blocks;
      ++blockIdx )
    {
      AccOmp5 ctx ( workDiv, blockIdx ); // OpenMP backend handle |\label{lst:accOmp}|
#     pragma omp parallel
      {
        functor ( ctx, args... );
      }
    }
  }

  // OpenACC
# pragma acc parallel |\label{lst:accParallel}|
  {
#   pragma acc loop gang
    for ( int blockIdx = 0;
      blockIdx < workDiv.blocks;
      ++blockIdx )
    {
      CtxBlockOacc ctxBlock ( workDiv, blockIdx ); // OpenACC backend handle |\label{lst:accOacc}|
#     pragma acc loop worker
      for ( int threadIdx = 0;
        threadIdx < workDiv.threads;
        ++threadIdx )
      {
         // need to add threadIdx to the context info
         AccOacc ctx ( ctxBlock, threadIdx );
         functor ( ctx, args... );
      }
    }
  }
}
\end{lstlisting}
 \caption{\label{fig:taskKernel}%
  Sketch of the structure of a \texttt{TaskKernel} template which calls a
  user-provided functor with user-provided argument in a parallel context using
  OpenMP or OpenACC, also providing a handle \texttt{ctx} to enable the user to
  generically access abstracted backend features. Note, that in the actual
  implementation \texttt{TaskKernel} is a class template which stores the
  functor and the arguments as members and provides a call operator to provide
  the functionality sketched above.
 }
\end{figure}

A block-shared variable \texttt{ctx} is declared inside the \texttt{teams},
respectively \texttt{loop gang}, region, but outside of the \texttt{omp
parallel}, respectively \texttt{loop worker} region, to represent a block-context
handle to the thread. This context information includes the current block id.
More details about this are provided in the following.

\subsubsection{Memory}

Alpaka's implementation adheres to the \gls{raii} principle and provides a
buffer type for memory on each device, including the host. Therefore, alpaka does not
have a concept of fixed associations between host and device memory, but rather
allows copying data between (sections of) any pair of buffers. One implication
of this is, that the buffer types for OpenMP and OpenACC must be implemented
using the manual management routines listed in table~\ref{tab:memAPI}, but more
importantly, that any pointer to data used in device code, will be a device
pointer with no associated host memory. This leads to different complications
with each of the models.

By default both models map local variables, including function parameters to the
device automatically if they are required in the device code. OpenMP assumes
that any pointer it maps hence contains a host address and tries to replace it by
the associated device address.
If no associated address exists the value copied to the device is \texttt{0x0},
which also applies to original device pointers.
The OpenMP runtime can be instructed to copy a device pointer verbatim
using the \texttt{is\_device\_ptr( \textit{varName}~)} clause on the
\texttt{target} directive in line~\ref{lst:ompTarget} or on additional
\texttt{omp target data} directives, which alpaka cannot use directly because
any pointers are elements of a parameter pack and thus have no name.
OpenMP will not perform this replacement on pointers which are enclosed in other
data types. Thus, we can get around this problem by wrapping any kernel
parameters, which are provided by the user in a C++ parameter pack, in an
\texttt{std::tuple} or similar structure.

OpenACC performs a similar replacement of address values also for pointers found
inside structures which are mapped, taking the wrapping-workaround off the table.
Like in OpenMP, a variable can be explicitly declared as a
device pointer using the \texttt{deviceptr( \textit{varName}~)} clause on the
\texttt{parallel} directive in line~\ref{lst:accParallel}, which is not possible
because there is no way to know which parameters contain pointers without access
to static reflection.
The only other option is to add
the clause \texttt{default(present)} telling the runtime to assume, that all
variables referenced in the offloaded code are already present on the device and
do not need to be mapped. Unfortunately this also disables automatic mapping of local
variables, necessitating that all local host variables used in the
\texttt{parallel} region occur in a \texttt{copyin} clause, which is feasible
because they are named.


\subsubsection{Block and Thread Index}

Each worker needs access to information about its position in the global
execution grid in order for work distribution to work. The OpenACC variant below
line~\ref{lst:accParallel} does basically show the canonical way of distributing
work over blocks and threads verbatim. In alpaka, the indices of these two
nested loops are passed to the user code in \texttt{ctx}. Because OpenMP
provides a build-in way to retrieve the thread index via
\texttt{omp\_get\_thread\_num()} we use a parallel region without loop to avoid
the overhead passing another loop counter to the user code, while the compiler
may use hardware intrinsics to supply the user code with this information.

\subsubsection{Atomic Operations}

The set of atomic operations supported by alpaka follows the set provided by
CUDA. This includes operations like \gls{cas}, min and max next to binary
operations. Both OpenMP and OpenACC support atomic load and store as well as
binary operations with both a pure \texttt{update} and a \texttt{capture}
semantic, i.e. atomically retrieving the stored value before applying the
operation to the memory location. Only OpenMP 5.1 adds an atomic
\texttt{compare} clause~\cite{openmp51atomic}, which permits the ternary
operator which is required to implement \gls{cas}, min and max.

As a work-around these ternary operations can be implemented using a
\texttt{critical} region, which, however, does not exist in OpenACC. Not
supporting these operations is not an option as PIConGPU's on-device dynamic
memory allocator mallocMC~\cite{mallocMC} requires them. Therefore we
had to implement critical region in violation of the OpenACC standard using
device-global, grid-level and block-level locks based on more basic atomics to
cover all levels at which atomic operations may be required.

\subsubsection{Block-Shared Memory}

Alpaka provides block-shared variables in the same way CUDA does, i.e.
declaration of them is allowed at any point inside kernel/thread code. This is
not supported by our targeted models. This capability is implemented by
providing a block-shared small-object allocator \texttt{BlockSharedSt} as part
of the \texttt{ctx} object, which contains a fixed-size member array as
underlying buffer. The size of this array can only be set at compile time.
Allocations of such shared variables are carried out by a master thread,
requiring synchronization between threads of a block, after which a reference to
the allocated memory is returned to all threads. A part of aforementioned buffer
can be reserved as a shared buffer of run time-size (\texttt{BlockSharedDyn}),
analogous to CUDA dynamic shared memory.

Our expectation is, that implementations will in time be optimized to actually
store block-shared variables which fit into on-chip memory there when targeting
GPUs. As long as these variables are stored in global device memory our strategy
will at least allow these variables to reside compactly in a cache close to
executing block.

OpenMP~5.0 added the \texttt{omp allocate} directive which allows explicit
allocation of memory through the OpenMP runtime, including an
implementation-defined allocator \texttt{omp\_pteam\_mem\_alloc} which may allow
the declaration of block-shared memory akin to using CUDA's
\texttt{\_\_shared\_\_} attribute. We did not test this option yet,
because functionally, albeit conceivably with lower performance, we can make due
with the implementation described above, which is required for OpenACC either
way, and using this feature could open another angle where incomplete OpenMP~5.0
compiler support may hit us.

\subsubsection{Block-Level Barrier}

Alpaka provides a block-level barrier akin to CUDA's \verb|__syncthreads()|
which can be implemented in OpenMP using the \texttt{omp barrier} directive.

OpenACC does not support any explicit synchronization, so we again violated the OpenACC standard
by implementing a barrier using atomic operations and spin loops on counters stored in
the block-shared context variable \texttt{ctx}. This implementation works on
GPUs in practice as long as the OpenACC runtime executes exactly the number of threads per
block the code expects it to.

\subsection{Final Touches: PIConGPU}

PIConGPU uses one global variable on the target device which is not abstracted
by alpaka and thus must be declared by an \texttt{acc declare
device\_resident( \textit{varName}~)}, respectively \texttt{omp declare
target( \textit{varName}~)}, directive in the PIConGPU code.

Global \texttt{constexpr} variables, used primarily as part of the simulation
definition, for the most part only exist at compile time and influence
what code is generated for offloading.
Constants whose value the compiler cannot optimize out do require explicit
mapping to the device. This is the case for
arrays which are dynamically indexed at run time and any other constant the
address of which taken in any context, e.g. if it is used as \texttt{this}
argument of a member function call.
We chose to not map constants without run time storage explicitly to limit our
changes to the application code outside of alpaka to a minimum.

\section{Major Hurdles and Discussion\label{s:issues}}
\subsection{Standards Issues}

\subsubsection{OpenACC/OpenMP: Static \texttt{constexpr} Mapping}

Both OpenACC and OpenMP require global variables to be marked explicitly for
availability in device code, making no exception for compile-time constants, so
formally all compile time constants have to be declared for the
target.\footnote{In~\cite{openmp50}: Section 2.10.4, Restrictions, bullet 5} In a
code like PIConGPU this would effectively lead to each simulation definition
file starting with \texttt{omp declare target} and ending with \texttt{omp end
declare target} or, worse, \texttt{acc declare copyin(
\textit{listOfAllConstexpr}~)}. While the OpenMP version of this only makes the
declare construct devoid of meaning, the OpenACC version actually hurts the code
maintainability because each addition or removal of a constant has to be
mirrored in a second place.

Most available implementations only raise warnings when a \texttt{constexpr}
variable is referenced in a target region without having been mapped because
there is no problem unless the compiler determines, that a \texttt{constexpr}
must be addressed at run time. In this case a reference to a symbol is
generated, but no definition, eventually leading to errors during dynamic linking
at run time.

For compile time constants, the standards should require symbols to be generated
for \texttt{constexpr} variables that are defined inline if needed, as it is
easier for the compiler to determine if a given \texttt{constexpr} require a
run time representation then for the programmer.  This would be more consistent
with inline functions, or functions defined within the same translation unit in general, not
requiring a \texttt{declare} to be callable from an offloaded region. It would also better
fit the C++-notion, that the compiler decides how to handle \texttt{constexpr}.

\subsubsection{OpenMP: \texttt{static constexpr} Members}

We started this work with only Open\-MP~4.5 being supported by compilers which
required types to not contain any \texttt{static} data members for them to be
considered mappable to target, with no exception for compile time constants.
Most compilers would only warn if mapped types contained \texttt{static} compile
time constants. However, GCC did throw an error here, which prevented us from
performing any testing of GCC's OpenMP or OpenACC\footnote{OpenACC does not
actually have this restriction, but GCC's implementation is based on the OpenMP
implementation and thus inherited this check.} implementations with PIConGPU
where, as in any \gls{tmp} code, \texttt{static constexpr} data members are too
commonplace to be removed as a workaround. OpenMP~5.0 solves this problem
by removing the restriction on \texttt{static} data members altogether.

\subsubsection{OpenACC: Lack of Explicit Block-Level Synchronization}

The lack of an explicit block-level thread barrier is the most
prominent issue our OpenACC port faces in terms of production-readiness.
Although we do have a functioning workaround for it, based on the standard not
allowing any assumption on thread scheduling, it is correct in stating, that our
solution using atomics and spin loops is not save. In practice this issue is
exacerbated by the standard API not providing any way to ascertain the actual
number of threads being run per block nor to force a certain number to be run,
leaving it to the user to find the maximum number of threads the runtime will
actually run per block on a given platform and instruct alpaka to not request
more than this. If the number actually running threads is smaller than expected,
our barrier implementation will dead-lock.

Any need for synchronization between threads could in theory be served using only
the implicit barrier after a \texttt{loop worker} region, but this is not
feasible in practice. On the surface this would require changing all kernels in
PIConGPU which employ barriers to employ loops to describe block-level
parallelism, circumventing the existing abstractions and introducing a second
implementation incompatible with, e.g., the CUDA backend. Even deeper structural
changes would be required where PIConGPU uses other libraries based on alpaka:
MallocMC provides an alpaka device function which returns information about the
allocator to all threads of a block using collective operations which require
synchronization inside the function.

When targeting any hardware which does actually not allow any assumptions about
threat execution but somehow still supports the implicit barrier after a
\texttt{loop worker} region, a compiler could still implement an explicit
barrier by first inlining the code surrounding the barrier and then
reordering it to split it into one region before the barrier and one after. Some
transformations required to achieve this may require assertions about the code
to be made which the base language does not allow in general, e.g. swapping
of inner and outer loops. Considering, that any code where data dependencies
block the required transformations would be using the barrier in an invalid way,
often leading to a dead-lock at run time, the compiler could raise an error when
it fails to transform the code to make the barrier implicit.

\subsubsection{C++: \texttt{std:tuple} Trivial Copy}

When it comes to offloading, it is very important, that types can be copied
bit-wise in order to transfer instances to the target. The C++ \gls{stl}
defines the type trait \texttt{std::is\_trivially\_copyable} which, if true,
guarantees that a bit-wise copy is safe for the given type. Composite types of
trivially copyable types are trivially copyable if the move and copy
constructors and assignment operators are trivial. For any type which is mapped
to target, clang checks this condition and issues a warning when it is not met.

The C++ \gls{stl} provides the type \texttt{std::tuple} which is very useful,
among other things, when storing parameter packs for later use, or to sneak
device pointers past the address mapping of the OpenMP runtime. Unfortunately,
the C++ standard does not require implementations of \texttt{std::tuple} to be
trivially copyable if all component types are. Both the GNU libstdc++ and LLVM's
libc++ implement \texttt{std::tuple} in a way that removes the
\texttt{std::is\_trivially\_copyable} trait. While they can still be copied
bit-wise in practice as long as all component types can be, this makes
\texttt{std::tuple} effectively unusable for offloading, because its use turns
an otherwise useful warning into noise. An active defect report against the C++
standard on this issue exists~\cite{cppTupleIssue}.

\subsection{Compiler and Runtime Issues}

PIConGPU is a rather large application, combing through almost all of the
C++ core language and, while alpaka only requires a subset of OpenMP or OpenACC
to implement its API, it probably happens to lean on aspects which are usually
not a focus when porting legacy applications, such as a wide range of atomic
operations, data sharing at block level and manual memory management.
Hence, the main roadblock to this work was, and remains, an immaturity of the
compilers around OpenMP \texttt{target} and OpenACC as the specifications are rather new.
Previously ported applications may use a common set of patterns, predominantly
straight parallel unrolling of loops over arrays, which is what
current compilers support best. When straying off the trodden path we find a
number of compiler bugs. With respect to C++ features especially, existing test
suites appear to contain mainly small examples each only using a very limited
set.

The compiler/runtime bugs can be grouped roughly into three categories: First
\glspl{ice} triggered in the compilers by the occurrence of, usually, more complex
C++ constructs, e.g. lambdas or parameter packs, occurring in or around target
regions. The second are \glspl{ice} in the backend, where either intermediate
representations from host and target appear to get mixed up or plain missing
features, like missing code generation for some atomics. If a code gets past these
two, there are run time errors triggered by faulty code generation or runtime
behavior.

The largest issue arising from all of these is that further development of
triggering code is blocked until the compiler gets fixed; unless a simple work
around is available, which turned out to be rare. Encountering a run time error
is especially tricky, as, chances are, there is only one compiler successfully
compiling an example for a specific target and thus there is no way to
independently test whether the example code itself it correct.

Having only one large application, PIConGPU, that has to iterate with the compiler
development and gets stuck at each encountered compiler bug would be too slow.
In our work, we were able to take advantage of alpaka's example codes and test
suite. With each small application only using a subset
of the backend's features, some of which may trigger different errors first,
we were able to report or debug those in parallel.


Close communication with compiler vendors also helps to alleviate the issue of
code not being testable due a lack of reliable compilers. With the very fresh
support of offloading, error messages raised by the compiler can often be
misleading and fail to point at the actual issue. In some cases an \gls{ice} can
be triggered by faulty user code, which is just handled badly inside the
compiler. It is likely, that invalid code is not often part of compiler test
suites.

\Glspl{ice} are rather simple to report and caused either by incomplete
implementations or some compiler internals which we cannot comment on. Run time
errors caused by compiler bugs require debugging on our part. Issues we
encountered that lead to wrong run time behavior of our program include an atomic
capture which falsely did not change the addressed value and variables declared at
block-level being shared between all blocks instead of being private to a single
block. Bugs leading to the runtime actively raising an error are simpler to
pin down, such as, e.g., a case where accessing an element of a struct member array which is
above a certain size in bytes would cause the runtime to report a ``GPU memory
error''.

\subsubsection{Compiler Error Messages}

Another problem encountered when debugging issues in C++ codes in general,
specifically when using vendor compilers backed by smaller development teams, is
unspecific error reporting.  The minor part of this can be unspecific messages
of the form ``invalid something'', which is almost useless information in the
absence of a very verbose context description. The major problem is a lack of
context information provided by non-mainstream compilers. There appears to be a
strong focus on a procedural style of programming, leading to the assumption,
that pointing to the line in the source code where an error occurs provides
sufficient information---in C++ it does not: When a code instantiates a
template, e.g., ten times, four of which contain something \emph{invalid},
printing the same line number four times is not helpful. The fact, that
\texttt{g++} and \texttt{clang++} compilers sometimes produce screens upon
screens of a single error message in some template instantiation is a source of
much ridicule of C++, but it is very important to have this information in any
non-trivial application. Template instantiation generates code at compile time.
Consequently, that one line may have four different meanings when the error
occurs and six more different meanings where the template is instantiated
without causing a compiler error.  Thus, it is important for the programmer to
know \emph{which} instance of the template causes the error, otherwise they are
left guessing.

We also observed bugs in the error reporting of compilers. For example, a
compiler may print an error message pointing to the correct line number, but
name the wrong file, specifically complaining about a line in a header, while
giving the name of the main source file.

\subsection{Preliminary Results\label{s:firstResults}}

Despite complete alpaka backends for OpenMP \texttt{target} and
OpenACC and explicit mappings of some bits of PIConGPU's code which
circumvented the abstraction layer, PIConGPU still cannot be successfully ported
to GPUs using OpenMP or OpenACC due to a number of open issues in
compilers and runtimes as highlighted in the previous section.
With Clang, we were able to use host as offload target for OpenMP. We
had the same option with NVHPC and OpenACC. We 
successfully ran PIConGPU via alpaka's OpenMP or OpenACC backend
using recent versions of the respective compiler on the host.

\paragraph{HelloWorld} is the most basic example for an alpaka application in
that is only runs one kernel which does nothing but print the thread and block
index of each thread. Successful execution shows, that the block-shared variables
in figure~\ref{fig:taskKernel}, lines~\ref{lst:accOmp} or~\ref{lst:accOacc}
could be created and that that they contain the correct block index for each
block. This works on GPU with OpenACC using NVHPC. There is one major bit of
complexity in this example though, in that it requires the runtime to provide
\texttt{printf()} on target. Clang's offloading runtime for AMD GPUs does not
provide this yet.

\paragraph{VectorAdd} performs a simple addition of two vectors filled with
random values on the device and checks the result on the host. This example
includes copying data between host and device in both directions, next to
executing a kernel containing a grid-strided loop. It is, however more portable
than \textit{helloWorld} in that is does not require any \texttt{c-lib} functions
in offloaded code. This example works successfully on GPU both with OpenACC
using NVHPC. With OpenMP targeting AMD GPUs this example works only with AMD's
ROC Clang version 4.3.0, while with clang's \texttt{main}
branch we see an issue where having large arrays as members in structs causes a
memory error, which makes starting any alpaka kernel fail.

\subsubsection{Alpaka test suite}

Targeting the host, all tests pass with OpenMP and
OpenACC using clang 11+ and NVHPC 21.7, respectively.

Using the most current upstream clang%
\footnote{Git commit
{c20cb5547ddd}}%
, most tests compile targeting AMD GPUs, some fail with \glspl{ice}, two due
to missing \texttt{cmath} symbols for device code. All tests fail at run time
when using clang \texttt{main} because of the aforementioned issue crashing any
alpaka kernel. Using ROC clang, the linker hangs with most of the tests.

With OpenACC, targeting NVIDIA GPUs, the majority of tests compile and succeed
at run time.  Most notably, tests of the block-level thread synchronization and
block-shared memory succeed on GPU.

\section{Conclusions and Outlook\label{s:concl}}

\glsresetall

This work attempted to map the feature set of the C++ \gls{tmp}-based
accelerator abstraction layer alpaka to the directive-based APIs OpenMP
\texttt{target}
and OpenACC. The main conceptual difficulty is that alpaka's API set is
heavily inspired by CUDA and thus requires some very specific capabilities to be
implemented, such as the possibility for user code to declare block-shared
variables anywhere instead of only up-front, which no existing API other than
CUDA and it's look-alikes supports. Here, alpaka could achieve better
performance portability by relaxing its adherence to CUDA and instead providing
a more general abstraction.

We described a number of problems with the present ecosystem encountered
introducing directive-based models into a \gls{tmp}-based and heavily abstracted
C++ code, which any other abstraction layer, such as Kokkos of RAJA, aiming to
support these models would have to face, too.
These issues highlight, that the main for OpenMP and OpenACC
use-case for these programming models is to add accelerator-offloading to legacy
code, which are usually large Fortran or C code-bases. These programming
languages mostly support a procedural programming paradigm and only offer a
limited options for abstraction beyond this. C++ allows very high levels of
abstraction using templates, which provides a quite rigorous formalism for code
generation and meta-programming. This lack of consideration for C++ \gls{tmp}
makes the standards treat compile time constants like any other (global)
variable. This seems reasonable in C or Fortran, where a compile time constant
is little more than a variable that is known to the compiler to not change a
runtime and that has a static life time. In C++, they can affect code generation
itself, leading to the expectation, that handling them is the responsibility of
the compiler. They may sometimes become a variable at runtime which does not
change, but mostly their storage is optimized out. A distinction should be made
between static life time and compile-time life time, where in the latter case
the compiler should be responsible for mapping any data still needed at run time.

OpenACC aims to distinguish itself from OpenMP by offering a less explicit way
describe parallelism in code. As in this work it was our goal map a very
explicit model onto it, we could not appreciate this. From our perspective, the main
distinction is OpenACC's strict adherence to data parallel principles. This is
interesting from an academic perspective in that it might enforce a cleaner
description of parallel code if fully embraced. When porting originally
sequential code to OpenACC it may not be much of a problem to follow this
path. We are, however, not aware of any hardware architecture in common use
exhibiting the restrictions enforced by OpenACC. Likewise, no other offloading
API enforces strict data parallelism, which causes existing parallel codes to be
build around established features such as explicit barriers and accessible worker
ids. Porting a code that relies on certain patterns to a programming model that
does not offer them results in substantial changes to the code and may end up
being more of a rewrite than a port, such as in the case of PIConGPU. In our
view, this makes OpenACC not a viable general-purpose parallel programming model
in its current form.

The primary issue we encountered was the immaturity of compilers both with
respect to the support for OpenMP \texttt{target} and OpenACC as well the interaction of
that support with C++ code. This forced us to follow a, nowadays, quite unusual
development approach, were we had write our code without the possibility to
validate it by compiling, because the compiler would fail at an \gls{ice}. In
order to progress we shifted our focus to compilers and targets where we saw the
fastest development with respect to the issues we encountered. For this reason,
we followed the clang mainline development on the OpenMP side. With OpenACC, we
were only able to make significant progress due to active support from the
NVHPC developers with finding and fixing bugs in their compiler and also testing
our code with their development compilers in between releases.

We continue to follow the compiler development on both the OpenMP and
OpenACC side to push towards improved compiler support and to address potential
issues in the backends presented here.

\subsubsection{Acknowledgments}
We want to thank Mathew Colgrove (NVIDIA) and the NVHPC team for help with
debugging both compiler and code issues, Ron Lieberman (AMD) for testing
PIConGPU with AOMP and advice on Clang in general and the SPEC High Performance
Group for testing and support.
We acknowledge the IT Center of RWTH Aachen for access to their
infrastructure and Jonas Hahnfeld for support.

This material is based upon work supported by the U.S. Department of
Energy, Office of science, and this research used resources of the Oak
Ridge Leadership Computing Facility at the Oak Ridge National
Laboratory, which is supported by the Office of Science of the
U.S. Department of Energy under Contract No. DE-AC05-00OR22725.

This work was partially funded by the Center of Advanced Systems Understanding
(CASUS) which is financed by Germany's Federal Ministry of Education and
Research (BMBF) and by the Saxon Ministry for Science, Culture and Tourism
(SMWK) with tax funds on the basis of the budget approved by the Saxon State
Parliament.

\bibliographystyle{splncs04}
\bibliography{article}

\end{document}